# A single-atom electron spin qubit in silicon

Jarryd J. Pla[1,2], Kuan Y. Tan[1,2†], Juan P. Dehollain[1,2], Wee H. Lim[1,2], John J. L. Morton[3], David N. Jamieson[1,4], Andrew S. Dzurak[1,2], Andrea Morello[1,2]

[1]*Centre of Excellence for Quantum Computation & Communication Technology.*

[2]*School of Electrical Engineering & Telecommunications, University of New South Wales, Sydney NSW 2052, Australia.*

[3]*Department of Materials, Oxford University, Oxford OX1 3PH, UK.*

[4]*School of Physics, University of Melbourne, Melbourne VIC 3100, Australia.*

[†] Present address: Department of Applied Physics/COMP, Aalto University, P.O. Box 13500, FI-00076 AALTO, Finland

**A single atom is the prototypical quantum system, and a natural candidate for a quantum bit – the elementary unit of a quantum computer. Atoms have been successfully used to store and process quantum information in electromagnetic traps[1], as well as in diamond through the use of the NV-center point defect[2]. Solid state electrical devices possess great potential to scale up such demonstrations from few-qubit control to larger scale quantum processors. In this direction, coherent control of spin qubits has been achieved in lithographically-defined double quantum dots in both GaAs[3-5] and Si[6]. However, it is a formidable challenge to combine the electrical measurement capabilities of engineered nanostructures with the benefits inherent to atomic spin qubits. Here we demonstrate the coherent manipulation of an individual electron spin qubit bound to a phosphorus donor atom in natural silicon, measured electrically via single-shot readout[7-9]. We use electron spin resonance to drive Rabi oscillations, while a Hahn echo pulse sequence reveals a spin coherence time ($T_2$) exceeding 200 µs. This figure is expected to become even longer in isotopically enriched $^{28}Si$ samples[10,11]. Together with the use of a device architecture[12] that is compatible with modern integrated circuit technology, these results indicate that the electron spin of a single phosphorus atom in silicon is an excellent platform on which to build a scalable quantum computer.**

There have been a number of proposals for the implementation of a spin-based qubit in silicon[13], though none have been studied in as much detail as the phosphorus atom qubit[14]. This interest has been motivated by the knowledge, developed over half a century from electron spin resonance experiments on bulk-doped phosphorus in silicon[15], that spin coherence times can be exceptionally long, exceeding seconds[11]. This is due to the availability of silicon in an enriched nuclear spin-zero ($^{28}Si$) form, as well as the low spin-orbit coupling in silicon[15]. The use of donor electron spins has further advantages of consistency (since each atom is identical) and tuneability (e.g. through the Stark shift[16]), while the donor atom's nuclear spin can be employed as a quantum memory for longer term storage[17].

Using methods compatible with existing complementary metal-oxide-semiconductor (CMOS) technology, we have fabricated a nanostructure device on the $SiO_2$ surface to enable read-out and control of an electron spin[12] (Fig. 1a). In this work, the donor is intentionally implanted into the silicon substrate, with future options including the use of deterministic ion implantation[18] or atomic precision in donor placement through scanning probe lithography[19]. The device is placed in a magnetic field of ~1 T, yielding well-defined electron spin-down and spin-up states ($|\downarrow\rangle$ and $|\uparrow\rangle$).

Transitions between the electron $|\downarrow\rangle$ and $|\uparrow\rangle$ states are driven by an ac magnetic field generated by applying microwaves to an on-chip broadband transmission line[4,20]. By operating at a high magnetic field and low temperature ($T_{electron} \approx 300$ mK), we can detect these transitions through single-shot projective measurements on the electron spin with a process known as spin-to-charge conversion[7,8]. Here the donor electron is both electrostatically and tunnel coupled to the island of a single electron transistor (SET), with the SET serving as both a sensitive charge detector and an electron reservoir for the donor. Using gates PL and TG (Fig. 1a) to tune the electrochemical potentials of the donor electron spin states ($\mu_\downarrow$ and $\mu_\uparrow$ for states $|\downarrow\rangle$ and $|\uparrow\rangle$) and the Fermi level in the SET island ($\mu_{SET}$), we can discriminate between a $|\downarrow\rangle$ or $|\uparrow\rangle$ electron as well as perform electrical initialisation of the qubit, following the procedure introduced in Ref. [8].

Our experiments use a two-step cyclical sequence of the donor potential, alternating between a spin readout/initialisation phase and a coherent control phase. The qubit is first initialised in the $|\downarrow\rangle$ state through spin-dependent loading by satisfying the condition $\mu_\downarrow < \mu_{SET} < \mu_\uparrow$ (Fig. 1b). Following this, the system is brought into a regime where the spin is a stable qubit ($\mu_\downarrow$, $\mu_\uparrow$ << $\mu_{SET}$) and manipulated with various microwave pulse schemes resonant with the spin transition (Fig. 1c). The spin is then read out electrically via spin-to-charge conversion (Fig. 1b), a process which produces a pulse in the current through the SET $I_{SET}$ if the electron was $|\uparrow\rangle$, and leaves the qubit initialised $|\downarrow\rangle$ for the next cycle.

The electron spin resonance frequency can be extracted from the spin Hamiltonian describing this system (see also Fig. 1d):

$$H = \gamma_e B_0 S_z - \gamma_n B_0 I_z + A\mathbf{S}\cdot\mathbf{I} \qquad (1)$$

where $\gamma_{e(n)}$ is the gyromagnetic ratio of the electron (nucleus), $B_0$ is the externally applied magnetic field, **S** (**I**) is the electron (nuclear) spin operator with z-component $S_z$ ($I_z$) and $A$ is the hyperfine constant. If $\gamma_e B_0 >> A$, the states shown in Fig. 1d are good approximations for the eigenstates of Eq. (1). Allowed transitions involving flips of the electron spin only (identified by arrows in Fig. 1d) exhibit resonance frequencies that depend on the state of the $^{31}$P nuclear spin: $\nu_{e1} \approx \gamma_e B_0 - A/2$ for nuclear spin-down; and $\nu_{e2} \approx \gamma_e B_0 + A/2$ for nuclear spin-up. The transition frequencies $\nu_{e1}$ and $\nu_{e2}$ are found by conducting an electron spin resonance (ESR) experiment[21], which is described in the Supplementary Information.

To demonstrate coherent control, we apply a single microwave pulse of varying duration $t_p$ to perform Rabi oscillations of the electron spin. For each $t_p$ the cyclic pulse sequence (Figs. 1e,f) is repeated 20,000 times, first with a microwave frequency $\nu_{e1}$, and immediately after at $\nu_{e2}$. It is necessary to pulse on both ESR transitions as the $^{31}$P nuclear spin can flip several times during acquisition of the data in Fig. 2a. Fig. 1g displays single-shot traces of the SET output current $I_{SET}$ for four consecutive repetitions of the measurement sequence, for an arbitrary pulse length. A threshold detection method[8] is used to determine the fraction of shots that contain a $|\uparrow\rangle$ electron for the measurements at both frequencies. Fig. 2a shows the electron spin-up fraction $f_\uparrow$ as a function of the microwave pulse duration for different applied powers, $P_{ESR}$. The fits through the data are derived from simulations assuming Gaussian fluctuations of the local field (see Supplementary Information). Confirmation that these are Rabi oscillations comes from the linear dependence of the Rabi frequency with the applied microwave amplitude ($P_{ESR}^{1/2}$), i.e. $f_{rabi} = \gamma_e B_1$. Here $B_1$ is taken as half of the total linear oscillating magnetic field amplitude generated by the transmission line at the site of the donor, assuming the rotating wave approximation. Fig. 2b shows the expected linear behaviour with microwave amplitude of the Rabi frequencies extracted from the data in Fig. 2a. The largest Rabi frequency attained was 3.3 MHz ($B_1 \approx 0.12$ mT), corresponding to a $\pi/2$ rotation in $\sim$ 75 ns.

The qubit manipulation time should be contrasted with the coherence lifetime of the qubit, termed $T_2$. Possible sources of decoherence include spectral diffusion of the $^{29}$Si bath spins[15,22,23], noise in the external magnetic field, and paramagnetic defects and charge traps at the Si/SiO$_2$ interface[24]. These mechanisms can, to a degree, be compensated for by utilising spin echo techniques (Fig. 3a), as long as the fluctuations are slow compared with the electron spin manipulation time (typically $\sim$ 100 ns).

Fig. 3a presents the gate voltage and microwave pulsing scheme for a Hahn echo measurement. Dephasing resulting from static local contributions to the total effective field during an initial period $\tau_1$ is (partially or fully) refocused by a $\pi$ rotation followed by a second period $\tau_2$ (see Fig. 3c for a Bloch sphere state evolution). A spin echo is observed by varying the delay $\tau_2$ and recording the spin-up fraction. In Fig. 3e we plot the difference in delay times ($\tau_2$ - $\tau_1$) against $f_\uparrow$. For $\tau_1 = \tau_2$, we expect to recover a $|\downarrow\rangle$ electron at the end of the sequence if little dephasing occurs (i.e. for short $\tau$), and hence observe a minimum in $f_\uparrow$. When $\tau_2$ - $\tau_1 \neq 0$, imperfect refocusing results in an increase in the recovered spin-up fraction. The echo shape is approximated as being Gaussian and the half-width at half-maximum implies a $T_2^* = 55 \pm 5$ ns.

We now set $\tau = \tau_1 = \tau_2$ and monitor the spin-up fraction as a function of $\tau$, to obtain the spin echo decay curve of Fig. 3f. A fit of the form $y = y_0 \exp(-(2\tau/T_2)^b)$, where $y_0$, $T_2$ and $b$ are free parameters, yields $T_2 = 206 \pm 12$ μs and $b = 2.1 \pm 0.4$. The coherence time $T_2$ is almost a factor of 2000 times longer than $T_2^*$, and is remarkably close to the value (300 μs) measured in bulk-doped natural silicon samples[25]. Variations in $T_2$ can be expected, depending on the exact distribution of $^{29}$Si nuclei within the extent of the donor electron wavefunction. This indicates that the presence of a nearby SET and the close proximity of the Si/SiO$_2$ interface have little, if any, effect on the electron spin coherence. This is not entirely surprising, since paramagnetic centres at the Si/SiO$_2$ interface are expected to be fully spin polarised under our experimental conditions $g\mu_B B_0 >> k_B T$ (where $g$ is the donor electron Landé g-factor, $\mu_B$ is the Bohr magneton and $k_B$ is the Boltzmann constant), leading to an exponential suppression of their spin fluctuations[26]. Direct flip-flop transitions between the donor qubit and nearby interface traps are suppressed by the difference in g-factor ($g = 1.9985$ for the donor, $g > 2$ for the traps[21]), whereas dipolar flip-flops with nearby donors[27] can appear as a $T_1$ process[8] on a much longer timescale. We measured $T_1 \approx 0.7$ s at $B_0 = 2.5$ T (data not shown), implying that this process has no bearing on $T_2$. The echo decay is Gaussian in shape ($b = 2.1 \pm 0.4$), consistent with decoherence dominated by $^{29}$Si spectral diffusion[22].

We have extended the coherence time by applying an XYXY dynamical decoupling ESR pulse sequence[28] (Figs. 3b,d). This sequence substitutes the single $\pi$ rotation of the Hahn echo with a series of four $\pi$ rotations alternating about the X and Y axis, achieved by applying adjacent $\pi$ pulses that are 90° out of phase. The resulting echo decay is shown in Fig. 3f, with a fit to the data yielding $T_2 = 410 \pm 20$ μs and $b = 2.1 \pm 0.4$. As well as representing a factor of 2 improvement in $T_2$, the XYXY sequence demonstrates the ability to perform controlled rotations about two orthogonal axes on the Bloch sphere (X and Y), permitting arbitrary one-qubit gates for universal quantum computing[29].

Next we consider the fidelity of our electron spin qubit, broken down into three components: measurement, initialisation and control. The measurement fidelity $F_M$ comprises errors resulting from detection limitations of the experimental setup as well as thermally induced readout events. The electrical spin-down and spin-up read errors ($\gamma_\downarrow$ and $\gamma_\uparrow$ respectively) arise from a finite measurement bandwidth and signal-to-noise ratio. They depend on the threshold current $I_T$ used for detecting the spin-up pulses. Fig. 4a shows the results of a numerical model based on our experimental data (see Supplementary Information for details), where $\gamma_{\downarrow,\uparrow}$ are plotted as a function of $I_T$. At $I_T = 370$ pA we achieve a best case error of $\gamma = \gamma_\downarrow + \gamma_\uparrow = 18\%$.

Thermal broadening of the Fermi distribution in the SET island produces the read/load errors, as depicted in Fig. 4b. The process of a spin-down electron tunnelling into an empty state in the SET occurs with a probability $\alpha$, whereas $\beta$ denotes the probability of incorrectly initialising the qubit in the spin-up state. The parameters $\alpha$ and $\beta$ are sensitive to the device tuning and can vary slightly between measurements. We have extracted $\alpha$ and $\beta$ from simulations of the Rabi oscillations in Fig. 2a, and for $P_{ESR} = 10$ dBm we find $\alpha = 28 \pm 1\%$ and $\beta = 1^{+9}_{-1}\%$. This gives an average measurement fidelity for the electron spin-up and spin-down states of $F_M = 1 - (\gamma + \alpha(1 - \gamma_\downarrow))/2 = 77 \pm 2\%$ and an initialisation fidelity $F_I$ of $\geq 90\%$ (see Supplementary Information for full details).

The qubit control fidelity $F_C$ is reduced by random field fluctuations from the $^{29}$Si nuclear bath spins. These produce an effective field $B_{eff}$ in the rotating frame that is tilted out of the XY-plane (Fig. 4d), and lead to imperfect pulses. We now estimate the strength of these fluctuations. Fig. 4c presents a series of ESR spectra, where the electron spin-up fraction is monitored as a function of the microwave frequency. The top three traces of Fig 4c contain individual sweeps with each point obtained over a timescale of $\sim 250$ ms. We attribute the shift in peak position between sweeps to slow fluctuations of a few strongly coupled $^{29}$Si nuclei, with hyperfine coupling strengths on the order of $\sim$ 1 MHz. The width of the peaks is most likely the result of distant, weakly coupled $^{29}$Si nuclear spins that fluctuate on the single-shot timescale (see Supplementary Information for further discussion). The bottom trace of Fig. 4c contains an average of 100 sweeps, representing many nuclear spin configurations. From this we extract a full-width at half-maximum $\Delta\nu = 7.5 \pm 0.5$ MHz. This is consistent with the observed $T_2^*$, where $\Delta\nu = 1/(\pi T_2^*) = 6 \pm 1$ MHz. To calculate the rotation angle error, we simulate a Rabi experiment assuming the largest $B_1$ achieved (0.12 mT) and Gaussian fluctuations of the nuclear bath with a standard deviation of $\sigma = \Delta\nu / \left(2\sqrt{2\ln(2)}\right) = 3.2 \pm 0.2$ MHz (see Supplementary Information). From this we infer an average tip angle of 102

± 3° for an intended π rotation, corresponding to an average control fidelity of $F_C = 57 \pm 2\%$.

The processes that contribute to the measurement, initialisation and control fidelity degradation can be mitigated with foreseeable adjustments to the device architecture and experimental setup. Significant improvements in the read/load errors would follow from enhanced electrical filtering to lower the electron temperature, thus enabling the high readout fidelities (> 90%) already achieved[8]. Moving to an enriched [28]Si (nuclear spin-zero) substrate[10] would remove the primary source of rotation angle error, and allow access to the exceptional coherence times already demonstrated in bulk-doped samples[11].

Future experiments will focus on the coupling of two donor electron spin qubits through the exchange interaction[14], a key requirement in proposals for scalable quantum computing architectures in this system[30]. Taken together with the single-atom doping technologies[18,19] now demonstrated in silicon, the advances reported here open the way for a spin-based quantum computer utilising single atoms, as first envisaged by Kane[14] more than a decade ago.

## METHODS SUMMARY

**Device fabrication and experimental setup.** For information relating to the device fabrication and experimental setup, we refer the reader to the Supplementary Information.

**Simulated quadrature detection for $T_2$ measurements.** For each $\tau$ ($\tau = \tau_1 = \tau_2$ for the Hahn echo), the sequence of Fig. 3a (Fig. 3b) is repeated 30,000 times (75,000 times) for the Hahn echo (XYXY dynamical decoupling) measurement at both $\nu_{e1}$ and $\nu_{e2}$, and for X and Y phases of the final π/2 rotation. The resulting signal amplitude is given by $(f_\uparrow(\nu_{e1}, Y) - f_\uparrow(\nu_{e1}, X)) + (f_\uparrow(\nu_{e2}, Y) - f_\uparrow(\nu_{e2}, X))$, where $f_\uparrow(\nu_{e1}, Y)$ represents the electron spin-up fraction of the single-shot traces taken at $\nu_{e1}$ with a final π/2 pulse about the Y-axis etc. The data points in Fig. 3f have been re-normalised with the amplitudes and offsets extracted from free-exponent fits through the decays. A 30% reduction in signal amplitude was observed for the XYXY dynamical decoupling decay, relative to that of the Hahn echo.

**Supplementary Information** accompanies the paper on www.nature.com/nature.

**Acknowledgements** We thank R.P. Starrett, D. Barber, C.Y. Yang and R. Szymanski for technical assistance. We also thank A. Laucht for the Bloch sphere artwork and D. Reilly for comments on the manuscript. We acknowledge support from the Australian Research Council and the Australian National Fabrication Facility. This research was supported in part by the U.S. Army Research Office (W911NF-08-1-0527).

Correspondence should be addressed to J.J.P. (jarryd@unsw.edu.au), A.M. (a.morello@unsw.edu.au).

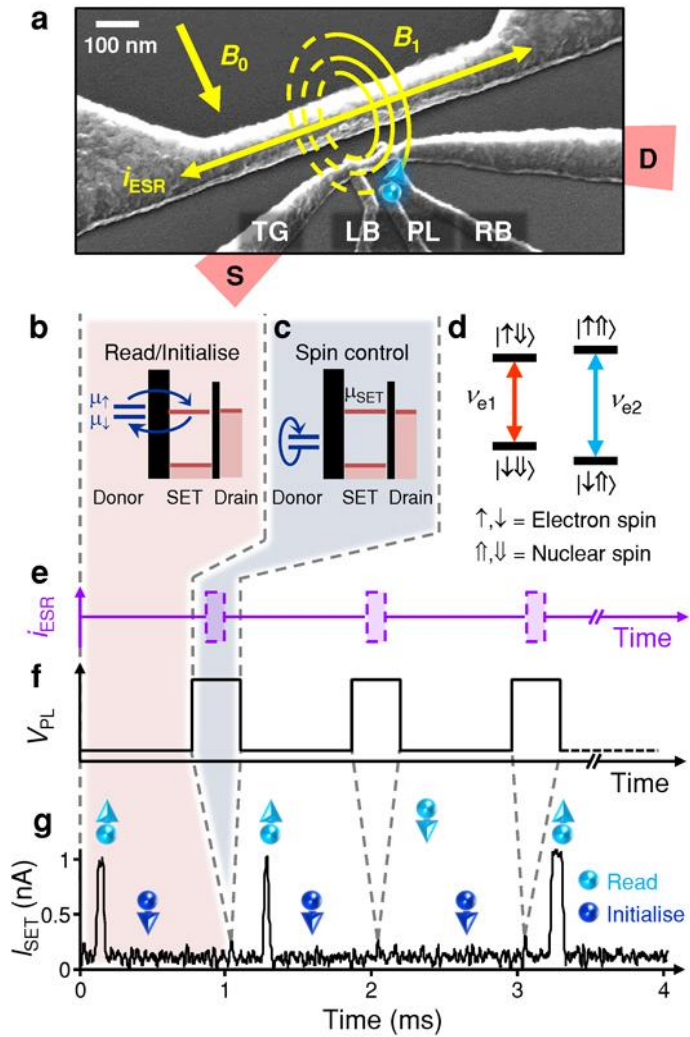

**Figure 1 | Qubit device and pulsing scheme. a,** Scanning electron micrograph of a qubit device similar to the one used in the experiment. The SET (lower right portion) consists of a top gate (TG), plunger gate (PL), left and right barrier gates (LB and RB) and source/drain contacts (S and D). The microwave transmission line is shown in the upper left portion. The donor (blue) is subject to an oscillating magnetic field $B_1$ from the transmission line which is perpendicular to the in-plane external field $B_0$. **b-c,** Pulse sequence for the qubit initialisation, control and readout. **b,** Read/initialisation phase $\mu_\downarrow < \mu_{SET} < \mu_\uparrow$: A spin-up electron will tunnel from the donor to the SET island, to later be replaced by a spin-down electron, causing a pulse of current through the SET. A spin-down electron remains trapped on the donor throughout the entire phase. **c,** Control phase $\mu_\downarrow$, $\mu_\uparrow << \mu_{SET}$: Electron spin states are plunged well below the SET island Fermi level whilst microwaves are applied to the transmission line to perform electron spin resonance. **d,** Energy level diagram of the $^{31}$P electron-nuclear system. **e-f,** Microwave pulse sequence (**e**) and synchronised PL gate voltage waveform (**f**) for performing and detecting spin manipulations (not drawn to scale). An arbitrary ESR pulse sequence is represented by the dashed purple box in panel **e**. **g,** Example of $I_{SET}$ response to four consecutive read/control events where a single microwave pulse of duration $t_p$ is applied, taken at $B_0$ = 1.07 T. The pulse duration $t_p$ has been set to give a high probability of flipping the electron spin. The duration of the pulses in $I_{SET}$ gives the electron spin-down tunnel-in time (~ 33 μs), whilst their delay from the beginning of the read phase gives the spin-up tunnel-out time (~ 295 μs).

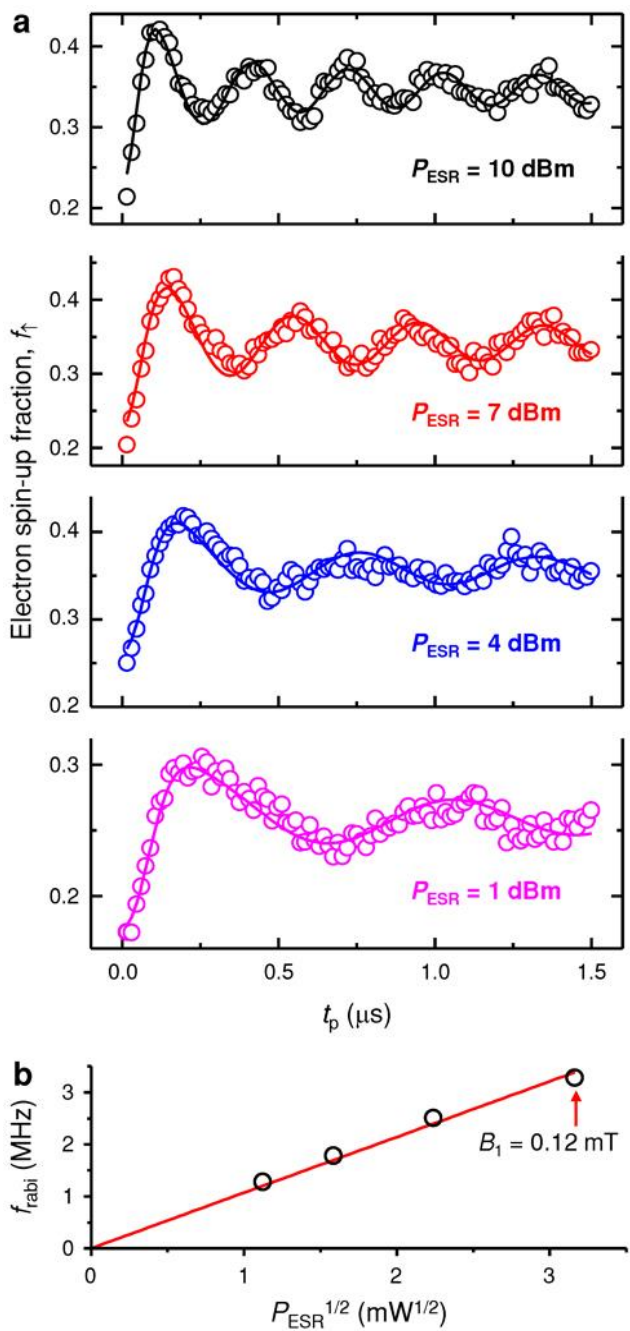

**a**

Electron spin-up fraction, $f_\uparrow$

$P_{ESR}$ = 10 dBm

$P_{ESR}$ = 7 dBm

$P_{ESR}$ = 4 dBm

$P_{ESR}$ = 1 dBm

$t_p$ (µs)

**b**

$f_{rabi}$ (MHz)

$B_1$ = 0.12 mT

$P_{ESR}^{1/2}$ (mW$^{1/2}$)

**Figure 2 | Rabi oscillations and power dependence of the Rabi frequency. a,** Electron spin-up fraction as a function of the microwave burst duration for varying input powers $P_{ESR}$. Measurements were performed at an external field of $B_0$ = 1.07 T where the ESR frequencies are $\nu_{e1}$ = 29.886 GHz and $\nu_{e2}$ = 30.000 GHz. Each point represents an average of 20,000 single-shot measurements, with each shot ≈ 1 ms in duration (see Supplementary Information for further details). The solid lines are fits generated from simulations of the measurements (Supplementary Information). **b,** Rabi frequency versus the microwave excitation amplitude, with a fit displaying the linear relationship.

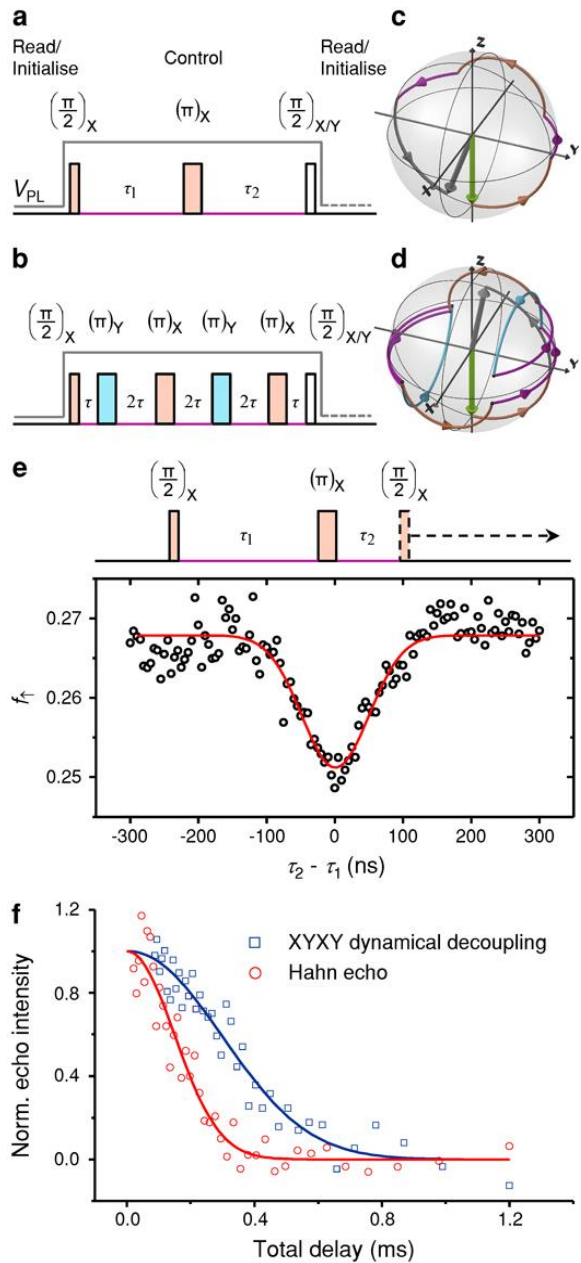

**Figure 3 | Coherence time and dynamical decoupling. a-b,** Pulse protocols for the Hahn echo (**a**) and XYXY dynamical decoupling (**b**) sequences with accompanying PL gate voltage waveforms, as described in the main text. The rotation angles are displayed above each pulse in brackets, with the subscript (X or Y) denoting the axis on the Bloch sphere about which the rotation is applied. The read/initialisation time is 1 ms. All measurements were performed at $B_0$ = 1.07 T and with $P_{ESR}$ = 10 dBm, where a $\pi/2$ rotation takes ~ 75 ns. **c-d,** Bloch sphere representation of the evolution in the rotating frame for the Hahn echo (**c**) and XYXY (**d**) sequences. The green arrow represents the initial spin state $|\downarrow\rangle$, whilst the grey arrow represents the final state for the case when the second $\pi/2$ pulse is about X (Y is not shown). The purple path represents dephasing in between pulses, the orange path represents a rotation about X, and the blue path is a rotation about Y. We have included rotation angle errors of 5° and 15° for the $\pi/2$ and $\pi$ pulses respectively. **e,** An echo curve, obtained by applying the depicted pulse sequence with a fixed $\tau_1$ (= 10 µs) and varying $\tau_2$. Each point represents the electron spin-up fraction $f_\uparrow$ calculated from 50,000 single-shots acquired at both ESR frequencies ($\nu_{e1}$ = 29.886 GHz and $\nu_{e2}$ = 30.000 GHz) and summed. The fit in red is Gaussian and of the form $f_\uparrow = B\exp(-[(\tau_2 - \tau_1)/C]^2) + D$. **f,** Hahn echo (XYXY dynamical decoupling) decay in red circles (blue squares), measured via simulated quadrature detection (see the methods section for details). A fit through the data is given by $y = y_0\exp(-(N\tau/T_2)^b)$, where $N = 2$ ($N = 8$) for the Hahn echo (XYXY dynamical decoupling) experiment. Parameter values are discussed in the main text.

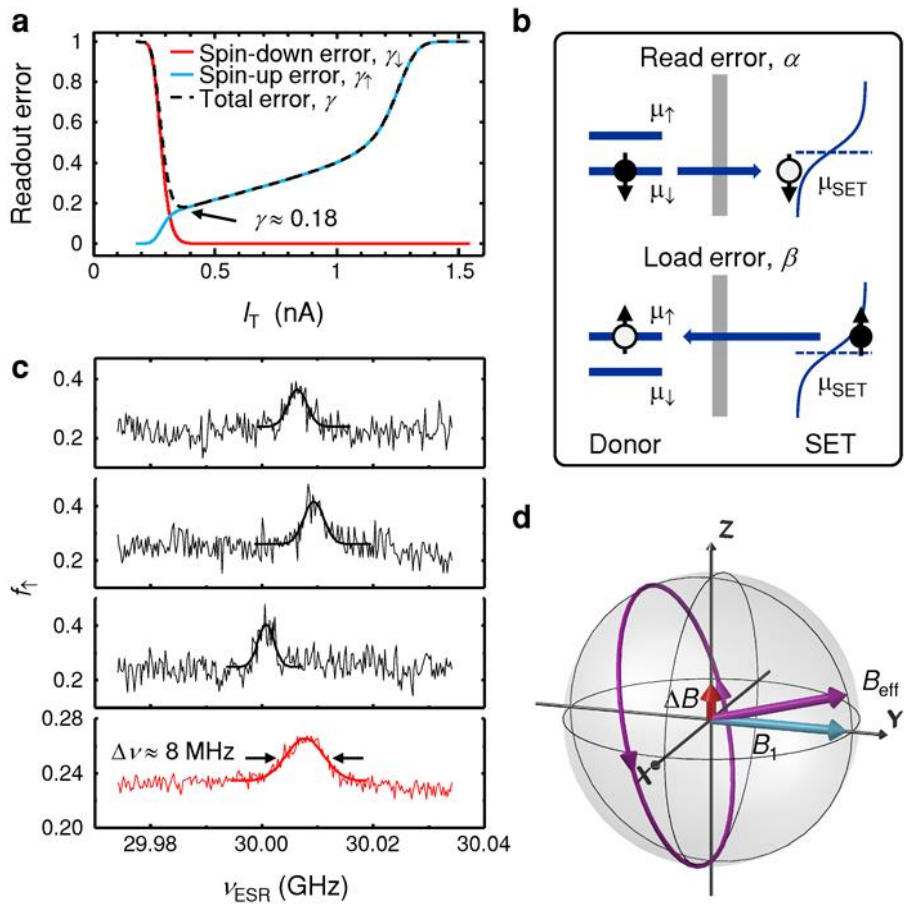

**Figure 4 | Qubit fidelity analysis. a,** Electrical readout errors generated from a numerical model. The red curve gives the error $\gamma_\downarrow$ involved in identifying a $|\downarrow\rangle$ electron as a function of the threshold current $I_T$, caused by noise in $I_{SET}$ exceeding $I_T$. The blue curve represents the error $\gamma_\uparrow$ for detecting a $|\uparrow\rangle$ electron, which occurs as a result of detection bandwidth limitations and a finite $|\uparrow\rangle$ $I_{SET}$ pulse height[8]. The dashed curve depicts the combined electrical error, $\gamma = \gamma_\downarrow + \gamma_\uparrow$. **b,** Mechanisms by which read (top) and load (bottom) errors are produced as a result of thermal broadening in the SET island (discussed in the main text). The solid circles represent full electron states with spin indicated by the arrow, whilst the empty circles signify unoccupied states. **c,** Sweeps of the frequency $\nu_{ESR}$ in the vicinity of the nuclear spin-up ESR transition $\nu_{e2}$. The top three traces are individual sweeps where $f_\uparrow$ at each $\nu_{ESR}$ is calculated from 250 single-shot measurements. The bottom trace is an average of 100 sweeps. **d,** Illustration of the rotation errors created by hyperfine field fluctuations of the $^{29}$Si nuclear bath. For simplicity, only the z-component of the hyperfine field has been shown. The bath nuclear spins produce an offset from resonance, $\Delta B$, which causes rotations about a new axis aligned with $B_{eff}$.